\documentclass[prl,superscriptaddress,twocolumn,nopacs,letter,10pt]{revtex4-2}
\usepackage{graphicx,bm,times}
\usepackage{amsmath}
\usepackage{amsfonts}
\usepackage{amssymb}
\usepackage{color}
\usepackage{hyperref}
\hypersetup{
	colorlinks = true,
	allcolors = {blue}
}

\usepackage[utf8]{inputenc}
\usepackage[T1]{fontenc}
\usepackage{amsmath,amssymb}
\usepackage{lineno}
\usepackage{float}
\usepackage{siunitx}
\usepackage{soul}
\usepackage{xfrac}

\newcommand{\rootthree}{$\sqrt{3}\!\times{}\!\sqrt{3}R(30^\circ{})$ }
\newcommand{\tworootthree}{$2\sqrt{3}\!\times{}\!2\sqrt{3}R(30^\circ{})$ }
\newcommand{\threebythree}{$3\!\times{}\!3$ }
\newcommand{\twobytwo}{$2\!\times{}\!2$ }

\begin{document}
\title{Change in charge density wave order beyond the Lifshitz transition in 2H-Ta\textsubscript{1$\pm\delta$}S\textsubscript{2}}

\author{Mihir Date}
\affiliation {Diamond Light Source Ltd, Harwell Science and Innovation Campus, Didcot, OX11 0DE, UK}
\affiliation {Max Planck Institute of Microstructure Physics, Weinberg 2, 06120 Halle (Saale), Germany}
\email{mihir.date@mpi-halle.mpg.de, \\matthew.watson@diamond.ac.uk}
\author{Jan Berges}
\affiliation {U Bremen Excellence Chair, Bremen Center for Computational Materials Science, and MAPEX Center for Materials and Processes, University of Bremen, D-28359 Bremen, Germany}

\author{Enrico Da Como}
\affiliation {Department of Physics, University of Bath, Claverton Down, Bath BA2 7AY, United Kingdom}

\author{Marcin Mucha-Kruczy\'{n}ski }
\affiliation {Department of Physics, University of Bath, Claverton Down, Bath BA2 7AY, United Kingdom}

\author{Alex Louat}
\affiliation {Diamond Light Source Ltd, Harwell Science and Innovation Campus, Didcot, OX11 0DE, UK}

\author{Gabriele Domaine}
\affiliation {Max Planck Institute of Microstructure Physics, Weinberg 2, 06120 Halle (Saale), Germany}

\author{Niels B. M. Schr\"{o}ter}
\affiliation {Max Planck Institute of Microstructure Physics, Weinberg 2, 06120 Halle (Saale), Germany}

\author{Malte Rösner}
\affiliation{Institute for Molecules and Materials, Radboud University, Heijendaalseweg 135, 6525AJ Nijmegen, The Netherlands}
\affiliation{Faculty of Physics, Bielefeld University, Universitätsstr. 25, 33501 Bielefeld, Germany}

\author{Matthew D. Watson}
\affiliation {Diamond Light Source Ltd, Harwell Science and Innovation Campus, Didcot, OX11 0DE, UK}

\begin{abstract}
We investigate electronic instabilities in 2H-TaS\textsubscript{2} and a self-intercalated variant, 2H$^\dagger$-Ta\textsubscript{1+$\delta$}S\textsubscript{2}. In conventional samples, which we determine to be slightly hole-doped, spectral gaps and backfolded features are found as fingerprints of the $3\times3$ charge density wave (CDW). Notably, the backfolded features emerge only at a temperatures below $T\approx$~65~K, substantially lower than the established CDW temperature of 78~K, suggesting an incommensurate-commensurate lock-in transition analogous to the phenomenology of the 2H-TaSe\textsubscript{2}. In contrast, the self-intercalated 2H$^\dagger$ sample exhibits substantial electron doping and signatures of a novel \tworootthree CDW. Using \textit{ab initio} calculations of the phonon spectrum, we demonstrate that the \threebythree instability ($\mathbf{q}=\sfrac{2}{3}\mathbf{\Gamma M}$) is highly sensitive to band filling. Furthermore, with increased interlayer spacing, a competing soft phonon mode emerges near $\mathbf{q}=\sfrac{1}{2}\mathbf{\Gamma K}$, corresponding to the superstructure observed in the 2H$^\dagger$ phase, although in our calculations this instability arises under hole doping rather than the electron doping inferred experimentally. These results establish band filling and interlayer spacing as key control parameters for CDW ordering vectors in 2H-TaS\textsubscript{2}, and highlight a route to engineering electronic instabilities in a prototypical layered material.
\end{abstract}

\date{\today}

\maketitle

Tuning through different many-body ground states as a function of band occupation is a hallmark of correlated electron systems. Metallic transition metal dichalcogenides (TMDs) provide a particularly fertile platform in this regard, hosting intertwined charge density wave (CDW), and superconducting order that can be tuned through pressure, strain, doping, and chemical substitution \cite{wilson_charge-density_1975}. The advent of artificial heterostructures has further expanded this landscape, enabling interface-driven modifications of their electronic structure and collective states \cite{vano_artificial_2021}. However, there is increasing recognition that the actual band filling may not reflect the expectation based on the nominal sample stoichiometry \cite{laverock_k-resolved_2013,straub_nature_2024}. Subtle adjustments of the band filling can drive Lifshitz transitions, reshape the Fermi surface, and profoundly influence both superconductivity and competing CDW instabilities \cite{hall_environmental_2019,luckin_controlling_2022}. 

Bulk 2H-TaS\textsubscript{2} is one of the canonical CDW systems. The literature establishes a \threebythree CDW transition temperature $\approx$~75-78~K \cite{naito_electrical_1982,wilson_charge-density_1975}.  Superconductivity has been reported with a $T_c$ of $\approx$1 K \cite{guillamon_chiral_2011}, but with a dome-shaped enhancement to over 4~K when the CDW is suppressed by doping \cite{wagner_tuning_2008,ni_crystal_2023}. However, the extent of CDW suppression depends on the nature of doping. For example, alkali and transition metal intercalation of 2H-TaX$_2$ (X=S, Se) weakens the lattice instability via electron doping~\cite{WilsonDiSalvoMahajan1974, wilson_charge-density_1975} merely due to chemical pressure of the intercalant~\cite{Friend01011987}. In contrast, self-intercalation does not suppress the CDW order completely but stabilises other commensurate orders such as $2\times1$, $2\times2$, and the $\sqrt{3}\times\sqrt{3}$ orders~\cite{zhao_engineering_2020, TianTian2023, Yunhui2025}.
\begin{figure*}[ht]
	\centering
        \includegraphics[angle=-90 , width=530pt]{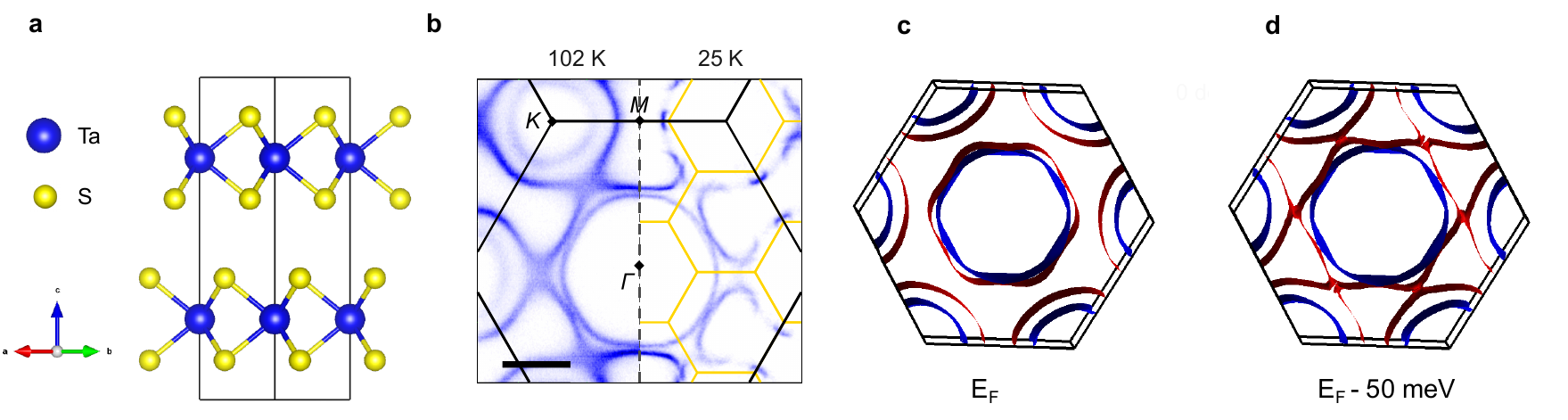}
	\caption{\textbf{Fermi surface topology in 2H-TaS$_2$ from experiment and DFT calculations.} \textbf{(a)} Crystal structure of 2Ha-TaS\textsubscript{2}. \textbf{(b)} Comparison of measured Fermi surfaces at 100~K and 25~K, above and well below the \threebythree transition temperature, with the normal (black) and reconstructed (orange) Brillouin zones overlaid ($h\nu{}$ = 58 eV, integrating spectral weight within 5 meV of $E_F$). Fermi surfaces calculated at \textbf{(c)} E$_F$, and \textbf{(d)} for a constant energy 50 meV below the Fermi energy. The latter shows the dogbone topology as in the experiments. Scale bars represent 0.5 \AA$^{-1}$. }
	\label{fig1}
\end{figure*}
The experimental hallmarks of lattice instabilities leading to the \threebythree order in the nominally stoichiometric 2H-TaS$_2$ are known from neutron X-ray scattering and Raman spectroscopy measurements~\cite{moncton_neutron_1977, SUGAI1981, ShenNatComms2023}. However, the selenide analogue 2H-TaSe\textsubscript{2} has been much more widely studied, including by photoemission measurements which exhibit clear signatures of momentum-dependent spectral gaps \cite{rossnagel_fermi_2005, borisenko_pseudogap_2008,laverock_k-resolved_2013,li_folded_2018}. Among the various reasons, the main factors limiting photoemission studies of TaS\textsubscript{2} have been structural heterogeneity and flakiness of the single crystals. Nevertheless, with the growing accessibility of microfocused angle-resolved photoemission spectroscopy ($\mu$-ARPES) the spectral imprints of the normal state and \threebythree ordered 2H-TaS$_2$ are starting to emerge~\cite{pudelko_probing_2024, camerano_darkness_2025}. An outstanding question in the sulphide case is the commensurability; while earlier literature pointed to a persistently incommensurate CDW \cite{scholz_charge_1982}, lacking the lock-in transition seen in the selenide \cite{tymoshenko_charge-density-wave_2025}, the topographic images accompanied by the local density of state measurements in scanning tunneling microscopy (STM) point to commensurate \threebythree superstructures in 2H-TaS$_2$ at low temperature~\cite{guillamon_chiral_2011, galvis_zero-bias_2014}. 

Here, using $\mu$-ARPES we reveal clear signatures of the known \threebythree CDW order on conventional 2H-TaS\textsubscript{2} samples, identifying small triangular bands from the Fermi surface reconstruction as a key signature. Interestingly, backfolded bands associated with commensurate \threebythree order are found only below $T\approxeq{}$~65~K, a phenomenology which suggests a hitherto unrecognised lock-in transition at a temperature below the established 78~K onset of the CDW. We further characterize a self-doped Ta\textsubscript{1+$\delta$}S\textsubscript{2} phase, found within a crystal of predominantly 4H$_b$-TaS\textsubscript{2}, with shrunken Fermi surfaces corresponding to a higher band filling of 1.16 electrons per unit cell, compared with 0.97 for the conventional samples. Intriguingly, the ARPES data indicates that a \tworootthree order is stabilised, qualitatively differing from the \threebythree in the band folding and gap structure. To investigate the potential microscopic origin of this, we complement our experimental results with \textit{ab initio} calculations that allow us to incorporate changes to the doping efficiently, demonstrating tunability of the phonon instabilities. The elongation of the $c$-axis, mimicking the reduced interlayer hopping found experimentally in the self-intercalated phase, is found to give rise to a softening of the phonon dispersion at $\mathbf{q}=\sfrac{1}{2}\mathbf{\Gamma K}$ that corresponds to the \tworootthree order. In our calculations, however, this instability appears for hole-doped 2H-TaS$_2$ instead of electron-doping as indicated by our data. Nevertheless, for both the normal and elongated calculations, the phonon instabilities are found to be highly sensitive to the band filling. Taken together, the results emphasize the importance and potential of using either natural or artificial control of the band filling to control the instabilities in this canonical quasi-2D metal. 

\section{Results and discussion}
Fig.~\ref{fig1}(a) shows the crystal structure of 2H-TaS$_2$ consisting of two trigonal prismatically stacked layers which are alternately rotated by 180$^\circ$, with the Ta atoms directly on top of each other, following the 2H$_a$ stacking \cite{katzke_phase_2004}. The Fermi surface measured in the high temperature undistorted phase at 102 K (left half of Fig.~\ref{fig1}(b)) contains 3 closed quasi-2D Fermi surfaces: a central ``circular" pocket around $\mathbf{\Gamma}$, twofold symmetric “dogbone” pockets centered at $\mathbf{M}$, and a single closed “barrel” around the $\mathbf{K}$-points. Contrastingly, at a temperature of 25~K, the $\mathbf{K}$-barrel is completely gapped, and the dogbone pockets develop partial spectral gaps, due to the formation of the \threebythree CDW \cite{borisenko_pseudogap_2008}.

This normal state Fermi surface topology, although familiar from the literature on bulk 2H-TaSe\textsubscript{2} \cite{rossnagel_fermi_2005,johannes_fermi_2008,laverock_k-resolved_2013}, differs qualitatively from the Fermi surface calculated by density functional theory (DFT; see Methods). As shown in Fig.~\ref{fig1}(c), DFT instead predicts two closed Fermi surfaces around each of the $\mathbf{\Gamma}$ and $\mathbf{K}$ points. Instead, the calculated constant energy map taken 50 meV below the Fermi level in Fig.~\ref{fig4}(d) reproduces the dogbone topology qualitatively well. 

The change in Fermi surface connectivity with band filling is known as a Lifshitz transition, which in this case is linked to whether a saddle point in the electronic structure found near $\sfrac{1}{2}\mathbf{\Gamma K}$ is above or below $E_F$ \cite{luckin_controlling_2022}. While the observed "dogbone" band topology implies hole-doping in our samples compared with the filling of exactly one electron per Ta in the calculation, the data is in agreement with other ARPES reports on 2H-TaS\textsubscript{2} \cite{camerano_darkness_2025, pudelko_probing_2024}, which also report the dogbone topology. The data in Fig.~\ref{fig1} can be therefore be considered to correspond to conventional 2H-TaS\textsubscript{2} samples. 

\begin{figure*}[ht]
	\centering
	\includegraphics[width=285pt, angle=-90]{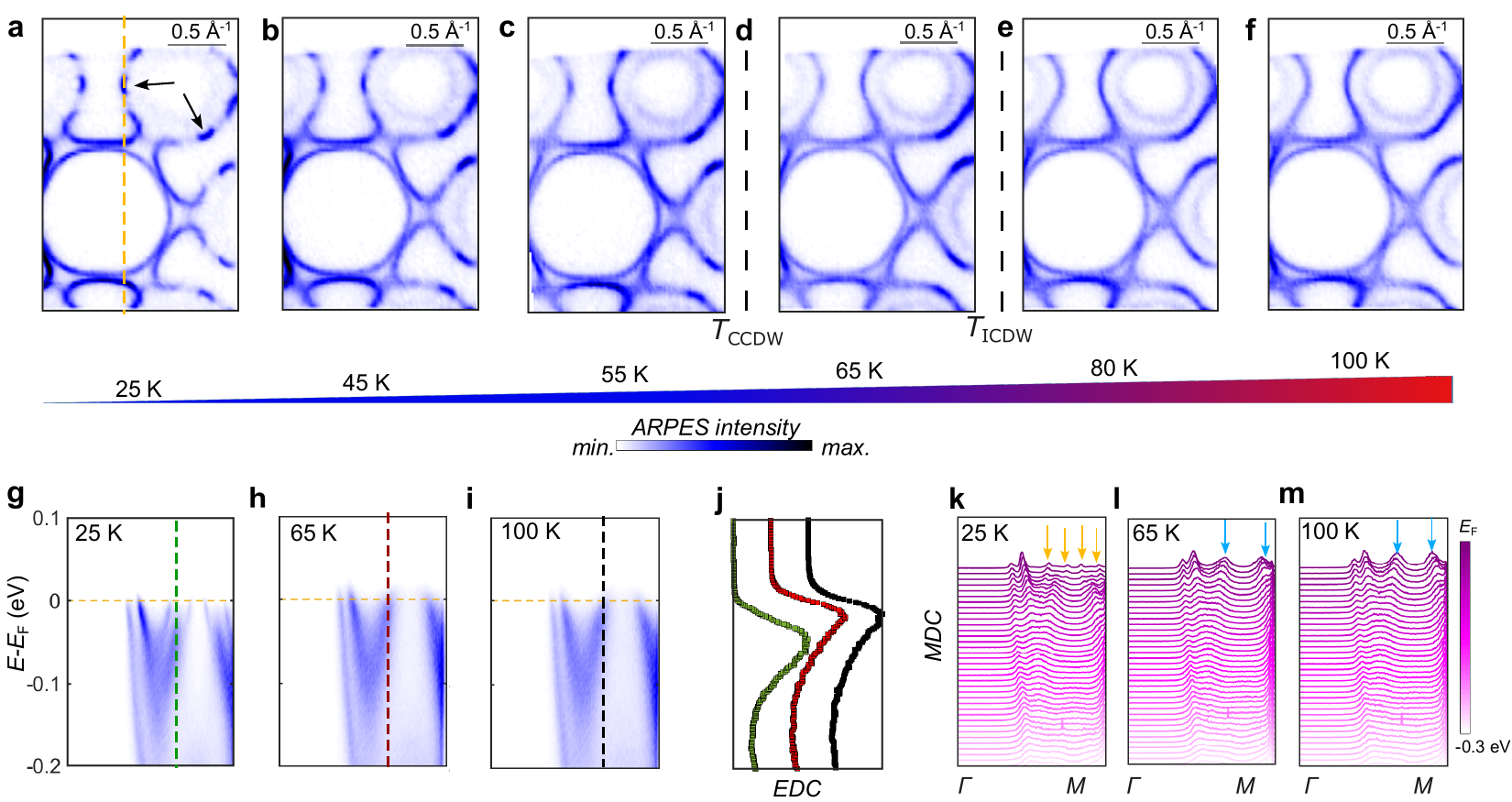}
	\caption{\textbf{Temperature-dependent ARPES on 2H-TaS\textsubscript{2}.} \textbf{(a)-(f)} Fermi surfaces of 2H-TaS\textsubscript{2} measured at $h\nu= 58$ eV in the temperature range of 25 - 100 K, showing the incommensurate CDW transition at approximately 78 K ($T_\mathrm{ICDW}$), and the lock-in transition around 55 K ($T_\mathrm{CCDW}$). \textbf{(g)-(i)} Dispersion along the direction marked by the dashed orange line in \textbf{(a)}, at 25, 65, and 100 K respectively. (j) EDCs at a constant momentum. \textbf{(k,l,m)} Corresponding MDCs near E$_F$ are displayed in\textbf{ (k)},\textbf{ (l)}, and \textbf{(m)}. There is no discernible difference between the 65 K and 100 K data. }
	\label{fig2}
\end{figure*}

The temperature-dependent Fermi surfaces shown in Fig.\ref{fig2}(a-f) offer deeper insights into the spectral properties of the 3$\times$3 phase conventional 2H-TaS$_2$. At low temperatures, the \threebythree order leaves clear fingerprints in the Fermi surface. First, the dogbone pocket becomes partially gapped. Second, along the $\mathbf{K}-\mathbf{M}-\mathbf{K}$ line the dogbone is not gapped but instead reconstructs into very small ``triangular" pockets with ``half-moon" like intensity as indicated by black arrows in Fig.~\ref{fig2}(a). The center of these triangular pockets corresponds to the equivalent of the $\mathbf{K}$-point in the \threebythree mini Brillouin zones, and their construction can be understood by considering backfolding of sections of Fermi surface by the wavevector $\sfrac{2}{3}\mathbf{\Gamma M}$ ~(see Supplementary Figure 1). These triangles are thus a hallmark of \threebythree order. Thirdly, the inner barrel pocket around the $\mathbf{K}$-point exhibits no spectral weight at all, appearing to be completely gapped. 

The most striking feature of our measurements in Fig.~\ref{fig2}, however, is not the low-temperature data, but rather the intermediate temperature regime. The data at 65~K, though nominally in the CDW phase, is spectrally much more similar to the normal state. Instead, the main spectral signatures of the 3$\times$3 phase become only visually apparent in the Fermi surface at $T$~<~55~K, with partially gapped dogbones and vanishing spectral weight on the inner $\mathbf{K}$-barrel. 

To understand this discrepancy further, we show the temperature evolution of the spectral function at a constant $k_x$ in Fig.~\ref{fig2}(g-i). We chose this cut, indicated by the yellow dashed lines, to intersect the dogbone pocket in two distinct places, revealing both spectral backfolding and the CDW gap. From the energy distribution curves (EDCs) shown in Fig.2(j), we estimated the CDW gap of the dogbone section to be approximately $\Delta~\approx$~35~meV at 25~K, whereas we found no such gap for the spectrum at 65~K. Moreover, the momentum distribution curves (MDCs) in Fig. 2 (k-m) show that the backfolded intensity peaks (shown in orange arrows) in the low-temperature spectrum vanish in the spectrum at 65 K, rather exhibiting an identical intensity profile to that of the MDC at 100 K. We have reproduced this temperature-dependence on another ARPES system, eliminating instrumental temperature calibration as a possible explanation.

The mismatch in the transition temperature of 75-78~K reported in the literature and the apparent onset temperature in ARPES demands explanation. Intriguingly, the phenomenology is reminiscent of 2H-TaSe$_2$ where there are known to be two transitions: a higher temperature transition at 122 K from a normal state to a slightly incommensurate 3$\times$3, followed by the so-called lock-in transition at 90 K. The former exhibits sharper anomalies in the heat capacity and resistivity ~\cite{harper_thermal_1977}, whereas the lock-in transition is most clearly seen in diffraction studies~\cite{moncton_neutron_1977,tymoshenko_charge-density-wave_2025}. Most pertinently, Borisenko et al.~\cite{borisenko_pseudogap_2008} pointed out that the main electronic reconstruction features as seen by ARPES occur only below the lower 90 K lock-in transition.\\

Returning to the case of 2H-TaS$_2$ in question here, while the CDW is usually stated as remaining incommensurate, the experimental evidence is sparse and mixed; a persistently incommensurate scenario was proposed based on electron diffraction~\cite{scholz_charge_1982,scholz_electron_1982-1}, but other techniques found it to be at least locally commensurate at low temperatures~\cite{nishihara_nmr_1983, guillamon_chiral_2011, wang_real-space_2022}, which is also implied by the observation of quantum oscillations~\cite{hillenius_quantum_1978}. A possible interpretation of our data, therefore, would be to support a previously unreported lock-in type phase transition between 55 and 65 K in 2H-TaS$_2$ analogous to the sister compound 2H-TaSe$_2$, becoming at least locally commensurate.

\begin{figure}[t]
    \centering
    \includegraphics[width=255pt]{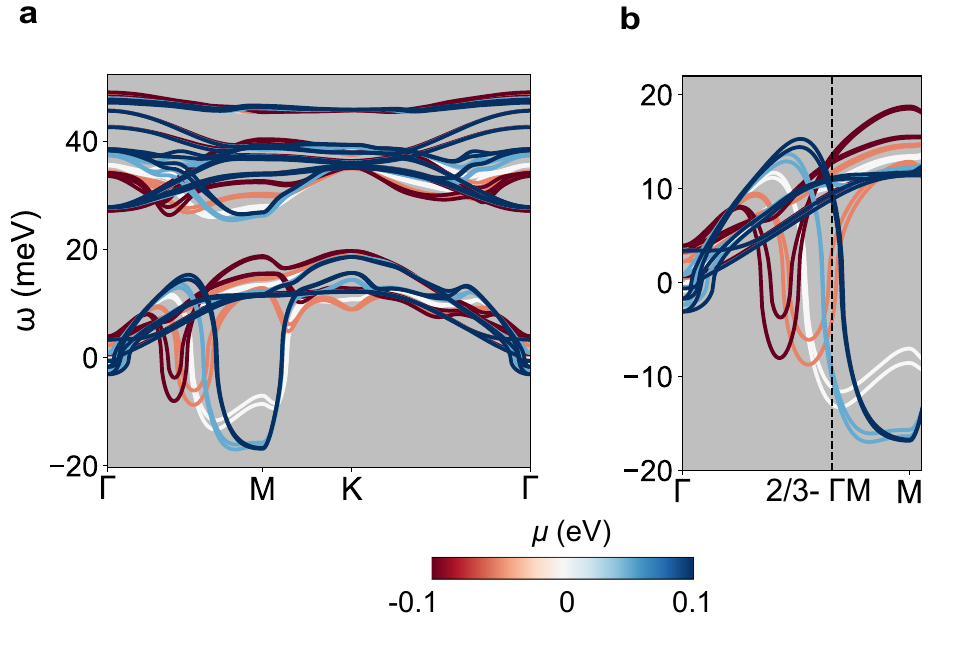}
    \caption{\textbf{Phonon bandstructures for regularly stacked 2H-TaS$_2$ with an interlayer spacing of 3.01 {\AA}}. \textbf{(a)} With changing doping levels, the longitudinal acoustic phonon mode softens at different points along $\mathbf{\Gamma-M}$. \textbf{(b)} Phonon dispersion zoomed in the $\mathbf{\Gamma-M}$ direction. The LA phonons in this region soften between $\sfrac{1}{2}\mathbf{\Gamma M}$ and $\mathbf{M}$ as the chemical potential is tuned from electron to hole doping.}
    \label{fig3}
\end{figure}

The drop in both resistivity and magnetic susceptibility at 78~K reported in the literature \cite{craven_specific_1977}, and also confirmed on our samples (Supplementary Figure 2), implies that there should be some loss of carrier density, or gapping of states at $E_F$, already at $T_{\mathrm{CDW}}$. Interestingly, careful inspection of the temperature dependent dispersion along the $\mathbf{M-K}$ direction shows signs of a partially gapped inner $\mathrm{K}$-barrel at 65 K (see Supplementary Figure 3) whereas the reconstruction in the dogbone appears only at 55 K and below. A comparison between these dispersions with the ones shown in Fig.\ref{fig2}(g-i) points towards two kinds of phase transitions: one at 78 K, and the other between 55 and 65~K. The occurrence of two transition temperatures with distinct spectral signatures further cements the scenario of a incommensurate to commensurate transition in 2H-TaS\textsubscript{2}. Since the incommensurate-commensurate transition in the selenide is usually understood in terms of free energy arguments \cite{mcmillan_theory_1976,littlewood_theory_1982}, our assertion that the same phenomenology occurs in the sulfide seems reasonable, and can reconcile much of the literature, though a revisit of this question with high resolution diffraction would be welcome. 


The 3$\times$3 CDW is a 3$\mathbf{q}$ ordering with $\mathbf{q}=\sfrac{2}{3}\mathbf{\Gamma M}$, and indeed our calculations of the phonon band dispersions for stoichiometric 2H-TaS\textsubscript{2} in Fig.~\ref{fig3} show a softening at approximately this wavevector. Motivated by the intrinsic doping in our sample, we studied the behaviour of the longitudinal acoustic (LA) phonons for different doping levels (see Methods) by shifting the chemical potential in the positive (electron doping) and negative (hole doping) directions. When hole doped (orange curves), the leading instability along the $\mathbf{\Gamma{}-M}$ direction persists, but the wavevector moves from $\sfrac{2}{3}\mathbf{\Gamma M}$ to approximately $\sfrac{1}{2}\mathbf{\Gamma M}$ where the latter corresponds to a 4$\times$4 charge order, which has not been observed. On the electron-doping side, the instability tends towards the $\mathbf{M}$ point, i.e. \twobytwo, reproducing earlier results \cite{hall_environmental_2019}. 

The variety and tunability of ground states that appear in the calculations motivates a search for possible alternative charge density wave ground states in TaS\textsubscript{2} samples at different doping. In samples nominally of the 4H$_b$-TaS$_2$ polytype, we found several inclusion regions in our spatial map where the stacking close to the surface was 2H-like. The occurrence of such stacking faults is common, and has been reported in X-ray diffraction and STM studies~\cite{Fujisawa2018STM, geng_correlated_2024}. Moreover, recent photoemission studies involving some of the authors of this work were able to identify regions having 2H-like stacking beneath a T-layer on the surface of 4H$_b$-TaS$_2$~\cite{Date2026}.

In what follows, we shall focus on this inclusion region with 2H-like stacking (which we label as 2H$^{\dagger}$) that shows a remarkably different CDW reconstruction, shown in Fig.~\ref{fig4} (a), compared to the regular 2H-TaS$_2$. First, we justify the assignment of the region based on the Ta 4\textit{f} core levels shown in Fig.\ref{fig4}(d). The bulk of this crystal was 4H$_b$-TaS\textsubscript{2}, in which case both H- and T- terminated regions can be observed \cite{ribak_chiral_2020, almoalem_charge_2024,watson_folded_2025,Date2026}. 4H$_b$-H terminations yield, in addition to the strongest doublet from the H termination, a  weaker split peak structure that is well-understood to arise from the Star of David reconstruction of the subsurface T layers \cite{hughes_site_1995,watson_folded_2025}. However, these are absent on the 2H$^\dagger$ region, implying an absence of T layers. Instead, only sharp peaks with binding energy corresponding to the trigonal prismatic (H) coordination are present. This leaves the possibility of 2H or 3R stacking, but given that there is a band splitting along the $\mathbf{\Gamma{}-M}$ direction in Fig.~\ref{fig4}(f), which is highly suppressed in the 3R stacking \cite{domaine_tunable_2025}, we assign this region as 2H stacking. We reproducibly found that crystals from this batch exhibited similar inclusion regions at the few percent level (Supplementary Figure 4). 

\begin{figure*}[ht]
	\centering
	\includegraphics[width=0.97\linewidth]{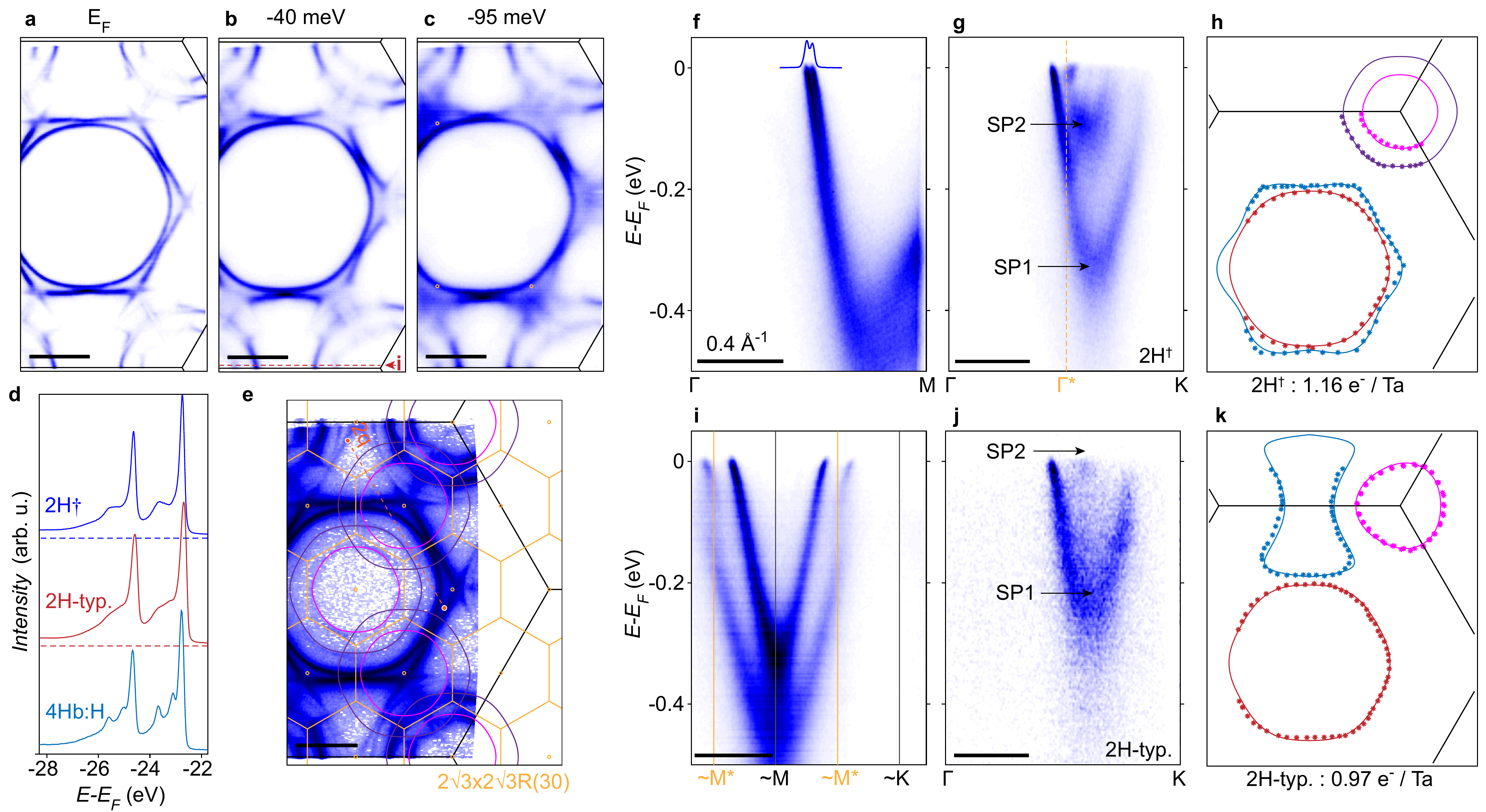}
	\caption{\textbf{Novel \tworootthree CDW in self-doped 2H$^\dagger$ Ta$_{1+\delta}$S\textsubscript{2}}. \textbf{(a)} Fermi surface and \textbf{(b,c)} constant energy contours at -40 and -95 meV respectively. \textbf{(d))} Comparison  of the Ta $4f$ shallow core level spectra. \textbf{(e)} Repeat of Fermi surface as in \textbf{(a)}, but with logarithmic color scaling to emphasize features with low spectral weight that derive from backfolding by the CDW. Orange lines represent the Brillouin zone of the \tworootthree expanded unit cell. Purple and magenta lines are guides to indicate backfolded features. Dashed line indicates backfolding of the hexagon and circle bands by twice the CDW wavevector. \textbf{(f,g)} High-symmetry cuts along the $\mathbf{\Gamma{}-M}$ and $\mathbf{\Gamma{}-K}$ directions. \textbf{(i)} Cut as marked by dashed line in \textbf{(b)}, showing hybridisation at the boundaries of the \tworootthree Brillouin zone. \textbf{(j)} High-symmetry cut along $\mathbf{\Gamma-K}$ for a conventional 2H-TaS$_2$ (2H-typ) sample. Compared to \textbf{(g)}, \textbf{(j)} appears to be electron doped. The extent of electron doping is estimated from the Luttinger analysis of the 2H$^\dagger$ and conventional 2H-TaS$_2$ Fermi surfaces in \textbf{(h)} and \textbf{(k)}, respectively.}
	\label{fig4}
\end{figure*}

The electronic properties of this 2H$^\dagger$ region are quite unlike the 2H-TaS\textsubscript{2} data reported until now. It is apparent that the normal/undistorted state of the 2H$^\dagger$ phase would have a Fermi surface with two barrels around the $\mathbf{K}$ point. The "dogbone" topology is found only in constant energy maps well below the Fermi level, such as in the constant energy contour 95 meV below $E_F$ in Fig.~\ref{fig4}(c). The $\mathbf{\Gamma{}-K}$ cut shown in Fig.~\ref{fig4}(g) indicates that both saddle points are occupied in 2H$^\dagger$, compared with only one in with the conventional 2H sample (Fig.~\ref{fig4}(k)). This region is therefore clearly substantially electron-doped. To quantify this, in Fig.~\ref{fig4}(h,l) we extract the Fermi contours from the data, interpolating across the regions showing spectral gaps to access the effective band structure of an undistorted $1\times{}1$ phase. We fit these with symmetry-appropriate functional forms (Supplementary Figure 5) and integrate the areas for the Luttinger count. We find a filling of 1.16 $e^-$ per Ta for the 2H$^\dagger$, contrasting with 0.97 $e^-$ per Ta for the conventional 2H-TaS\textsubscript{2} samples. 

A hint at a chemical explanation for the electron doping comes from a second look at the Ta $4f$ core levels in Fig.~\ref{fig4}(d): in contrast to the broader double-shoulder seen in the conventional 2H region, for 2H$^\dagger{}$ there is instead a single minority peak. We suggest that this phenomenology could be due to a partial filling of the Ta interstitial site as the primary defect structures \cite{han_phase_2024}. The sharper main peaks, together with the extremely low background in the ARPES measurements, would align with this, since interstitial Ta would cause less in-plane disorder compared with in-plane vacancies or inclusions. A secondary effect of the interstitial filling may be to increase the $c$ axis separation, which in turn would reduce the interlayer hopping. Evidence for this comes from the much reduced splitting along the $\mathbf{\Gamma{}-M}$ direction at just 0.026 \AA$^{-1}$, compared with 0.06 \AA$^{-1}$ in conventional samples (note that this splitting goes to zero in a freestanding monolayer) as seen in Fig.~\ref{fig4}(f). 

Most interestingly, the data in Fig.~\ref{fig4} show clear effects of a CDW but with very different spectroscopic signatures compared to the  \threebythree order. Most prominently, in Fig.~\ref{fig4}(a) sections of both the inner and outer $\mathbf{K}$ barrels are strongly gapped, but other sections are ungapped. The inner circle band around $\mathbf{\Gamma}$ seems to be unaffected by the CDW, but the corners of the hexagon exhibit strong hybridisation gaps at a binding energy of 40 meV (Fig.~\ref{fig4}(b,g)). Furthermore, there are numerous backfolded features, highlighted using the logarithmic colour map in Fig.~\ref{fig4}(e). The brightest backfolded features correspond to copies of the $\mathbf{K}$ barrels but shifted by a wavevector of $\mathbf{q=\sfrac{1}{2}\Gamma K}$, as illustrated by the construction in Fig.~\ref{fig4}(e). We find some evidence that this weight is further scattered by $\mathbf{q=\sfrac{1}{2}\Gamma K}$, forming a weak spectral copy around the zone centre. There is also weak but discernible backfolding of the hexagon and circle states, shown by the dotted orange line; this corresponds to wavevector of $\mathbf{K}$, twice the primary wavevector. 

The data thus point towards a \tworootthree CDW, corresponding to a $3\mathbf{q}$ instability with $\mathbf{q= \sfrac{1}{2}\Gamma K}$. Perhaps the most convincing evidence for the \tworootthree order is the cut presented in Fig.~\ref{fig4}(i), close to $\mathbf{M-K}$ (see Fig.~\ref{fig4}(b)). This shows $\sim$100 meV energy gaps precisely where the outer $\mathbf{K}$ barrel crosses the BZ boundaries of the \tworootthree order. This order is different from all previous suggestions for variants of (Ta,Nb)(S,Se)\textsubscript{2}. The \twobytwo order ($\mathbf{q=M}$) was proposed by Luckin \textit{et al}~\cite{luckin_controlling_2022} for highly potassium-dosed TaSe\textsubscript{2}, where the FS topology similarly changes to the 2-barrel scenario \cite{luckin_controlling_2022}, however it cannot account for the data here. \rootthree ($\mathbf{q}=\mathbf{K}$) is a superstructure which is known in self-intercalated Ta\textsubscript{1+$\delta$}S\textsubscript{2} {\cite{wang_giant_2022} and intercalated NbS\textsubscript{2} \cite{edwards_chemical_2024}, and indeed we do have some weak backfolding by a wavevector of $\mathbf{K}$. However, the primary backfolding of the $\mathbf{K}$ barrels demands a shorter $\mathbf{q}$ value, and the backfolded weight of the circle and hexagon bands scattered by $\mathbf{K}$ can be considered a second order effect, or a harmonic of the CDW, consistent with its very weak spectral weight (around 2$\%$ of the intensity of the main bands). Thus, the only option that appears compatible with all of the data is \tworootthree. 

A \tworootthree superstructure has been encountered previously in the literature on metallic transition metal dichalcogenides, albeit sparsely. A superstructure consistent with \tworootthree is present in some regions of alkali-intercalated 2H-TaS\textsubscript{2} samples \cite{ohta_evaluation_2020}, and a \tworootthree superstructure has been encountered in Ag\textsubscript{$\approx$0.33}NbSe\textsubscript{2} \cite{mogami_appearance_2021,curzon_electron_1985,rajora_electrical_1986} as well as Mn\textsubscript{x}NbS\textsubscript{2} \cite{mushenok_two-step_2023} and K\textsubscript{0.4}MoS\textsubscript{2} \cite{bin_subhan_charge_2021}. It seems probable to us that the \tworootthree order in TaS\textsubscript{2} may have a relatively narrow stability range, although in Supplementary Note 4 we show another region with the same Fermi surface topology, albeit with a slightly higher electron count of 1.20 $e^-$/Ta, exhibiting similar spectral gaps.

\begin{figure}[t]
    \centering
    \includegraphics[width=255pt]{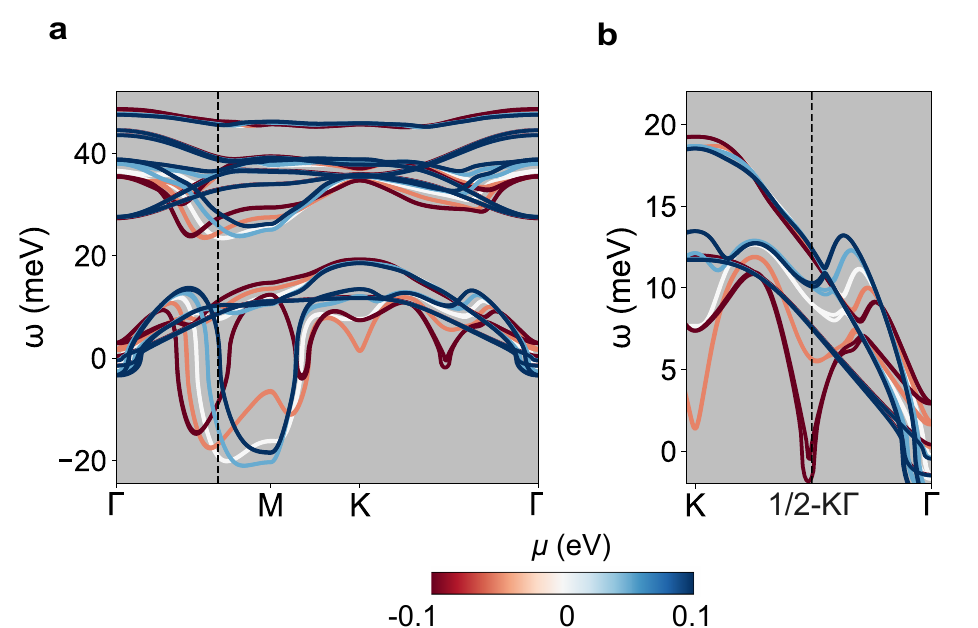}
    \caption{\textbf{ Phonon dispersion of 2H-TaS$_2$ with elongated \textit{c-}axis.}\textbf{(a)} For 2H-TaS$_2$ with the elongated \textit{c-}axis, the anomalies in the phonon bandstructure mainly mimic Fig.~\ref{fig3}(a), however, at $\mu=-50~\mathrm{meV}$, the LA phonons acquire negative frequency along $\mathbf{M-K}$. \textbf{(b)} Enlarged phonon dispersion in the $\mathbf{K-\Gamma}$ region shows soft LA phonons at $\sfrac{1}{2}\mathbf{\Gamma K}$.}
    \label{fig5}
\end{figure}

One of the intriguing features of a \tworootthree CDW is that the saddle points SP2 found along $\mathbf{\Gamma{}-K}$, as shown in Fig.~\ref{fig4}(g), all map close to the $\mathbf{\Gamma^*}$ point in the superlattice BZ (Fig.~\ref{fig4}(e)). Rice and Scott \cite{rice_new_1975} were the first to propose a CDW driven by saddle points in the electronic structure, an idea recently rekindled by Luckin \textit{et al} \cite{luckin_controlling_2022}. However, the evidence points against this mechanism playing a leading role here. Crucially, the SP2 saddle point is also quite deep in the occupied states, at -95 meV. Band hybridisation is seen in the vicinity of the saddle points, seen prominently in the constant energy map at -40 meV in Fig.~\ref{fig4}(b). However, at the Fermi level  in Fig.~\ref{fig4}(a), where the gaps are more relevant for the energetics, it is noticeable that only the very corners of the hexagon pocket are gapped, in comparison to large sections – approximately half – of the two barrels around the $\mathbf{K}$ point. As such, it is hard to make the case that the saddle point mechanism is dominating the energetics here. Instead, the CDW seems to be led by the partial gapping of the two \textbf{K} barrels. 


Noting that the \tworootthree CDW order arises experimentally in a system where self-intercalation may increase the interlayer spacing, we explore the phonon dispersion in 2H-TaS\textsubscript{2} with a slightly elongated \textit{c}-axis. Fig.~\ref{fig5} shows the resulting phonon dispersion for different band filling (i.e. different chemical potential values) along the high-symmetry directions. The trends in the electron-doped phonon dispersion shown in Fig.~\ref{fig5}(a) broadly mimic Fig.~\ref{fig3}(a), especially along $\mathbf{\Gamma{}-M}$. However, it is noteworthy that when hole-doped at $\mu=-0.05~\mathrm{eV}$, the anomaly along $\mathbf{M-K}$ becomes rather pronounced compared to Fig.~\ref{fig3}(a), and in fact the LA phonons almost soften at $\mathbf{K}$. Remarkably, the LA phonons develop additional anomalies along $\mathbf{M-K}$, and at $\sfrac{1}{2}\mathbf{\Gamma K}$ when hole-doped at $\mu=-0.1~\mathrm{eV}$ (see Fig.~\ref{fig5}(b)). The phonon softening at $\sfrac{1}{2}\mathbf{\Gamma K}$ occurs at an electron loss (hole-doping) of approximately 0.24 $\mathrm{e^-}$/Ta whereas, based on our Luttinger analysis, the \tworootthree order emerges at an electron gain (electron-doping) of approximately 0.16$\mathrm{e^-}$/Ta. While our experimental and theoretical calculations are therefore not fully reconciled, we have demonstrated that the charge orders in 2H-TaS$_2$ are sensitive to shifts in chemical potential, and interlayer spacing, with the latter having a larger impact on the acoustic phonon softening, and together with our results promote \tworootthree into the pantheon of orderings in 2H-TaS\textsubscript{2}.

\section{Summary and outlook}

To summarise, we have first shown that conventional 2H-TaS\textsubscript{2} with the familiar dog-bone Fermi surface is in fact slightly hole-doped compared to a hypothetical stoichiometric 2H-TaS\textsubscript{2}, and exhibits \threebythree CDW order. Our temperature dependent ARPES measurements present a compelling case for an incommensurate to commensurate (lock-in) CDW transition that is accompanied by spectral gaps in the $\mathbf{K}$-barrel and the dogbone pockets. Our \textit{ab initio} calculations align with this argument, with an instability found along $\mathbf{\Gamma{}-M}$ close to $\mathbf{q=\sfrac{2}{3}\Gamma M}$, although the leading $\mathbf{q}$ is strongly doping-dependent. In contrast, our measurements on self-intercalated and electron doped 2H$^\dagger$-TaS\textsubscript{2} reveal a previously unreported CDW, a \tworootthree phase corresponding to a $3\mathbf{q}$ order with $\mathbf{q_{CDW}}=\sfrac{1}{2}\mathbf{\Gamma K}$. We simulated this scenario by assuming 2H-TaS\textsubscript{2} with an elongated $c$-axis. In such a case, we do diagnose an instability at $\mathbf{q= \sfrac{1}{2}\Gamma K}$, although in the calculations it is stabilised with hole doping rather than the electron doping found in the experimental sample. The prediction of a CDW order along this wavevector has been found in previous works in the form of soft LA phonon modes~\cite{joshi_short-range_2019} and peaks in the electronic susceptibility across a wide doping range~\cite{luckin_controlling_2022} but has not received much attention since there was no experimental confirmation of such a phase until now. While the leading instability still lies along $\mathbf{\Gamma{}-M}$ in all cases, this is not necessarily a matter of concern because at low temperatures, the enhancement in the phonon self-energy at $\mathbf{q=\sfrac{1}{2}\mathbf{\Gamma}K}$ may suppress other instabilities in other regions of the Brillouin zone (see Ref.~\cite{Berges2020}), and for our system, we leave this for future inquiry. Nevertheless, we highlight that a new, and previously overlooked CDW order may be anticipated by modifying the interlayer spacing, and doping. Importantly, the \tworootthree order represents a class of CDW orders that emerge due to changes in the Fermi surface topology alongside with a sizeable electron-phonon interaction. While the idea of controlling many-body ground states in Ta dichalcogenides by tuning the band filling is not new – dating back at least to Wilson’s 1975 review \cite{wilson_charge-density_1975} - our experimental results and theoretical calculations add the previously overlooked \tworootthree order into the picture. 

\textbf{Methods}

\textbf{Experimental Methods.}
Samples of conventional TaS\textsubscript{2} were purchased from HQ graphene, who used a chemical vapor transport growth method. The crystals used to produce the datasets in Fig.~\ref{fig4} of the main text, and Supplementary Figure 3 were also grown by chemical vapour transport, with iodine as the transport agent. Stoichiometric amounts of tantalum and sulfur were loaded in quartz ampules and vacuum sealed. The 2H-TaS$_2$ crystals were grown in a two zone tube furnace with the hot/cold zones at 800$^\circ$ and 700$^\circ$C, respectively, for 12 days. The furnace was then cooled to room temperature at a rate of 0.1 C/minute. The procedures for the growth of 4H$_b$-TaS2 have been reported in Ref.~\cite{watson_folded_2025}. ARPES measurements were performed on the nano branch of the I05 beamline. The typical energy resolution was 20 meV. TaS\textsubscript{2} samples have notoriously "baklava"-like surfaces, especially after cleavage, with thin an often bent flakes appearing at the surface, necessitating the use of micro-ARPES combined with careful searching for areas exhibiting sharp features.

\textbf{Theory Methods.}
The first-principle calculations were performed using plane-wave density functional theory (DFT) as implemented in the QUANTUM ESPRESSO package~\cite{QE-2009, QE-2017}. The ion-electron interactions were described by fully-relativistic norm-conserving pseudopotentials compiled from the PSEUDO-DOJO library ~\cite{PseudoDojo2018}. The exchange-correlation functional was described by the Perfew-Becke-Erzenhoff (PBE) paramterisation of the generalised gradient approximation (GGA)~\cite{PBE1996}. We used a kinetic energy cutoff of 100 Ry, and a 12$\times$12$\times$4 Monkhorst-Pack~\cite{MonkhorstPack1976} \textit{k-}mesh for Brillouin zone integration, and a smearing of approximately 0.136 eV. To preserve the in-plane symmetries of 2H-TaS$_2$, the Ta positions were kept fixed, whereas only the z-coordinate of the S atoms was allowed to relax until all the Hellmann-Feynman forces on the atoms were less than 10$^{-5}$ Ry/Bohr.  The \emph{ab initio} phonon dispersion was then calculated over \textit{q-}grid of $6\times6\times2$ points.\\
A tight-binding model was then constructed using maximally localised Wannier functions projected over Ta-$d_{x^2-y^2}$, $d_{xy}$, and $d_{z^2}$ orbitals (active subspace) to capture the essential features of the DFT Fermi surface. The doping dependent renormalised phonon dispersions were calculated using the Calandra-Profeta-Mauri (CPM) scheme~\cite{CPM2010, CalandraMauriNbSe2} where the phonon self-energy $(\Pi_{\mathbf{q}\nu})$ obtained from density functional perturbation theory (DFPT) is patched by subtracting the phonon self-energy of the active subspace ($\Pi^\mathrm{A}_{\mathbf{q}\nu})$ calculated over a coarse \emph{k-}mesh, and subsequently adding $\Pi^\mathrm{A}_{\mathbf{q}\nu}$ calculated over a fine \emph{k-}mesh of 84$\times$84$\times$28 points, and the desired chemical potential $\mu$. 
The phonon self-energy calculations were performed using Eq.~\ref{phself} as implemented in the EPW package~\cite{EPW2016},
\begin{equation}
\Pi_{\mathbf{q}\nu}(\omega;T)
=
\frac{2}{N_k}
\sum_{mn\mathbf{k}}
\frac{
f_{n\mathbf{k}}(T)
-
f_{m\mathbf{k+q}}(T)
}{
\varepsilon_{n\mathbf{k}}
-
\varepsilon_{m\mathbf{k+q}}
}
\left| g_{mn\nu}(\mathbf{k},\mathbf{q}) \right|^2
\label{phself}
\end{equation}
where, $f_{n\mathbf{k}}$ (T) describes the Fermi-Dirac distribution at a \textit{k-}point for band \textit{n}, and $g_{mn\nu}(\mathbf{k,q})$ is the momentum-dependent electron-phonon matrix elements coupling electronic states indexed by \textit{n,m} with phonon modes $\nu$.




\section{References}
\bibliographystyle{naturemag}
\bibliography{ref_matt}

\textbf{Data Availability} 
The data that support the findings of this study are available from the corresponding author upon reasonable request. \\

\textbf{Acknowledgements}\\
We thank T.K. Kim, C.Cacho, and P.D.C. King for insightful discussions. We acknowledge Diamond Light Source for time on beamline I05 under proposal SI36633. J.B. would like to acknowledge funding from "Deutsche Forschungsgemeinschaft (DFG) under Germany's Excellence Strategy (University Allowance, EXC 2077, University of Bremen). 
\\\\
\textbf{Author Contributions}\\
M.D. and M.D.W. performed ARPES measurements with assistance from A.L., N.B.M.S, and G.D. E.D.C. grew the single-crystals of 2H-TaS$_2$ and 4H$_b$-TaS$_2$. M. D. performed \emph{ab intio} phonon calculations with help from J.B., and M.R. M.D. and M.D.W. wrote the paper with inputs from N.B.M.S., M.M.K, and all the other authors.  
\\

\textbf{Competing Interests Statement}\\
The authors declare no competing interests.

\end{document}